%% file: paper.tex
\documentclass[sigconf]{acmart}

\copyrightyear{2026}
\acmYear{2026}
\setcopyright{cc}
\setcctype{by}
\acmConference[CHIIR '26]{2026 ACM SIGIR Conference on Human Information Interaction and Retrieval}{March 22--26, 2026}{Seattle, WA, USA}
\acmBooktitle{2026 ACM SIGIR Conference on Human Information Interaction and Retrieval (CHIIR '26), March 22--26, 2026, Seattle, WA, USA}

\acmDOI{10.1145/3786304.3787870}
\acmISBN{979-8-4007-2414-5/2026/03}

\usepackage{tabularx}
\usepackage{subcaption}
\usepackage{graphicx}
\usepackage{booktabs}
\usepackage{todonotes}
\usepackage[section]{placeins}
\usepackage{comment}
\usepackage{enumitem}
\usepackage{multirow}
\usepackage{makecell}
\usepackage{balance}

\begin{document}
\title{Cooking Up Politeness in Human–AI Information Seeking Dialogue}
%\title{From Queries and Clicks to Conversations and Blends: How Users Engage with Web and Chat Systems}

    \author{David Elsweiler}
\affiliation{
  \institution{University of Regensburg}
  \city{Regensburg}
  \country{Germany}}
\email{david.elsweiler@ur.de}
\author{Christine Elsweiler}
\affiliation{
  \institution{University of Innsbruck}
  \city{Innsbruck}
  \country{Austria}}
\email{christine.elsweiler@googlemail.com}
\author{Anna Ziegner}
\affiliation{
  \institution{University of Innsbruck}
  \city{Innsbruck}
  \country{Austria}}
\email{Anna.Ziegner@student.uibk.ac.at}

\renewcommand{\shortauthors}{}

\input{00_abstract}

%%
%% The code below is generated by the tool at http://dl.acm.org/ccs.cfm.
%% Please copy and paste the code instead of the example below.
%%


%%
%% Keywords. The author(s) should pick words that accurately describe
%% the work being presented. Separate the keywords with commas.

%\received{20 February 2007}
%\received[revised]{12 March 2009}
%\received[accepted]{5 June 2009}

\maketitle

\input{01_introduction}
\input{02_related_work}
\input{03_frummet_analyses}
\input{04_simulation}

\input{06_discussion}
\bibliographystyle{ACM-Reference-Format}
\balance
\bibliography{bibliography}

\end{document}

%% file: 00_abstract.tex
\begin{abstract}
Politeness is a core dimension of human communication, yet its role in human–AI information seeking remains underexplored. We investigate how user politeness behaviour shapes conversational outcomes in a cooking-assistance setting. First, we annotated 30 dialogues, identifying four distinct user clusters ranging from \textit{Hyperpolite} to \textit{Hyperefficient}. We then scaled up to 18,000 simulated conversations across five politeness profiles (including impolite) and three open-weight models. Results show that politeness is not only cosmetic: it systematically affects response length, informational gain, and efficiency. Engagement-seeking prompts produced up to 90\% longer replies and 38\% more information nuggets than hyper-efficient prompts, but at markedly lower density. Impolite inputs yielded verbose but less efficient answers, with up to 48\% fewer nuggets per watt-hour compared to polite input. These findings highlight politeness as both a fairness and sustainability issue: conversational styles can advantage or disadvantage users, and ``polite'' requests may carry hidden energy costs. We discuss implications for inclusive and resource-aware design of information agents.

\end{abstract}

%% file: 01_introduction.tex
\section{Introduction}\label{sec:introduction}  

Generative AI (GenAI) systems are increasingly embedded in everyday communication, from customer service and online learning to workplace collaboration and creative writing. As their use expands, they not only shape human communication \cite{kouwenhoven2025shaping,tanguy-etal-2025-human,thomas2020expressions} but are also shaped by the language practices of their users \cite{wang2025under}. Optimised to produce coherent and broadly acceptable outputs \cite{leiter2024chatgpt}, these systems risk contributing to a homogenisation of language and a loss of linguistic diversity \cite{kuteeva2024diversity}. Since variation encodes cultural values, social identities, and interactional norms \cite{agha2003social, kiesling2013constructing}, it should be accounted for in human--GenAI exchanges.  

A central dimension of such variation is \textit{politeness}, i.e. ``how we maintain good relations and avoid interpersonal conflict through the use of different linguistic forms and strategies'' \cite{culpeper2017introduction}. Politeness norms differ across and within language communities \cite{schneider2017politeness, schneider2022pragmatic} and directly shape communication across different settings. In human-AI interactions, studies show that users value polite agents \cite{kumar2022politeness} but often adopt more direct styles themselves for efficiency \cite{kodur2025exploring, lumer2023indirect}. %With conversational agents,
While human input varies widely, agent responses remain comparatively uniform \cite{barko2020conversational}, influencing perceptions of quality. Although prompt politeness has been shown to affect model output \cite{yin2024should}, we still know little about how politeness unfolds in task-oriented GenAI dialogues where efficiency and clarity are paramount.  

At the same time, societal debates have framed politeness as inefficient, since it adds tokens that increase computational cost and energy use \cite{delavande2025thankyou,orf2024politeness}. Reducing interaction to machine-optimal forms, however, risks privileging technical constraints over human communicative needs and may disadvantage groups whose styles are underrepresented in training data. This makes politeness not only a question of usability but also of fairness and inclusion. Unlike well-studied demographic and content biases (e.g., gender \cite{kotek2023gender}, toxicity  \cite{ousidhoum2021probing}), pragmatic variation such as politeness has rarely been studied in task-oriented GenAI contexts.
%While bias in large language models (LLMs) has been studied in domains such as gender \cite{kotek2023gender} and toxicity \cite{ousidhoum2021probing}, the role of politeness variation remains largely unexplored.  

To address this gap, we focus on \textit{task-based information-seeking dialogues}, where users pursue concrete goals but social aspects of communication remain relevant. %Cooking assistance provides an exemplary testbed for structured, stepwise information seeking. By combining empirical analysis with controlled simulation, we show how politeness strategies shape efficiency and inclusivity in this domain, with implications for task-oriented GenAI interactions more broadly.
Cooking assistance provides an exemplary testbed: it combines structured, stepwise information needs with opportunities for socially oriented behaviours such as thanking, hedging, or apologising. We take a two-step approach. First, we analyse Frummet et al.’s dataset of Wizard-of-Oz cooking conversations \cite{frummet2024cooking} to examine how users employ politeness and engagement strategies when interacting with a passive assistant. Second, we extend these insights through LLM--LLM simulation, systematically manipulating user politeness profiles across agent models. Our study addresses three central questions:  
\begin{enumerate}
    \item \textbf{Response length (RQ1):} Do politeness profiles elicit longer or shorter responses?  
    \item \textbf{Information transfer (RQ2):} Do politeness profiles affect the amount of information conveyed per step?  
    \item \textbf{Energy efficiency (RQ3):} Do politeness profiles influence the efficiency of information transfer, linking pragmatic variation to debates on sustainable and fair GenAI?  
\end{enumerate}  

By combining empirical analysis of human data with controlled simulation, we show how politeness strategies impact efficiency and inclusivity in GenAI-mediated cooking tasks. Our findings provide both insight into pragmatic variation and practical guidance for designing conversational agents that balance efficiency with fairness. %   and cultural sensitivity.  %Can we really provide guidance on cultural sensitivity based on our study?

\begin{comment}
References to think about

Pragmatics / Politeness Theory

Brown, P., & Levinson, S. C. (1987). Politeness: Some universals in language usage. Cambridge University Press.

Leech, G. (2014). The pragmatics of politeness. Oxford University Press.

Blum-Kulka, S., House, J., & Kasper, G. (1989). Cross-cultural pragmatics: Requests and apologies. Ablex.

Bias in AI and Language Homogenisation

Bender, E. M., Gebru, T., McMillan-Major, A., & Shmitchell, S. (2021). On the Dangers of Stochastic Parrots: Can Language Models Be Too Big? �� Proceedings of FAccT. https://doi.org/10.1145/3442188.3445922

Blodgett, S. L., Barocas, S., Daumé III, H., & Wallach, H. (2020). Language (Technology) is Power: A Critical Survey of “Bias” in NLP. Proceedings of ACL. https://doi.org/10.18653/v1/2020.acl-main.485

Hovy, D., & Spruit, S. L. (2016). The social impact of natural language processing. Proceedings of ACL. https://doi.org/10.18653/v1/P16-2096

Cultural Communication & Technology

Haugh, M. (2018). The role of culture in interaction. In A. Capone et al. (Eds.), Pragmatics and its Interfaces. Springer.

Scollon, R., & Scollon, S. W. (2001). Intercultural communication: A discourse approach. Blackwell.
\end{comment}

%% file: 02_related_work.tex
\section{Related Work}\label{sec:related_work}

The related work is structured in two subsections. The first examines prior research on politeness within the field of pragmatics, while the second reviews studies on the linguistic dimensions of conversational information systems. Together, these discussions establish the foundation and rationale for our research agenda.

\subsection{Variation in linguistic politeness}

The systematic study of linguistic politeness in pragmatics, including the use of speech acts realising different politeness strategies, goes back to the seminal work of Lakoff, Brown \& Levinson and Leech in the 1970s and 80s \cite{lakoff1973logic, brown1987politeness, Leech1983principles}. Since then, it has been demonstrated that politeness norms not only vary across different linguacultures such as English and German \cite{blum-kulka_cross-cultural_1989, house2021cross} but also within the same language \cite{schneider_variational_2008, schneider2017politeness}. Intralingual variation is influenced by both demographic (e.g., region, age or gender) and situational factors (e.g., power, social distance or interactional setting) \cite{Barron_2021}. Age has, for instance, been shown to impact the level of directness in requests in German \cite{kranich2021requests}. Regional differences are evident in the use of \textit{please}, which is often avoided in American English, particularly when there is power imbalance, but is a routine politeness marker in British English \cite{murphy2019routine}. The impact of communicative setting is evident, e.g., in digital communication, as different digital formats such as online chats, text messaging or email communication engender distinct politeness conventions \cite{graham2017politeness, thaler2012sprachliche, merrison2012getting}. These may even vary across languages or national varieties of languages, as the realisation of email requests by British and Australian undergraduate students to faculty highlights \cite{merrison2012getting}. While the British students mark power differences through deference,  Australian students treat their lecturers as social peers by emphasising closeness and common ground. 

Politeness variation has important implications for human–GenAI interactions. If GenAI systems are more attuned to certain politeness norms than others, they may respond in ways that advantage some user groups while disadvantaging others, even within the same language community. Understanding politeness patterns is therefore critical to designing systems that can interpret and respond appropriately to diverse politeness styles.

\subsection{Language in Information Interaction}

The role of language in human–AI interaction has been studied from several complementary perspectives, with a focus on conversational style, system politeness, and user politeness.

Conversational style has been shown to influence the flow and perceived quality of interactions. Thomas et al. demonstrate that style can be systematically measured in spoken information-seeking dialogues, using features such as pronoun use, repetition, and speech rate \cite{thomas2020expressions}. Users were found to adapt their style toward the agent’s, while style matching by the agent improved conversational smoothness and reduced the negative effects of errors \cite{thomas2018style}. More recent work confirms that LLMs likewise adapt syntactically to their interlocutor, aligning their output with user style \cite{kandra2025llms}. The  style features adopted by LLMs, e.g., complexity or friendliness, moreover have an impact on user preferences, which are in turn influenced by user traits such as extroversion and agreeableness \cite{chevi2025individual}. These findings underscore the role of linguistic alignment in human–agent dialogue, but they centre primarily on style rather than politeness as a pragmatic dimension.

System politeness has mostly been studied from the perspective of user perceptions \cite{ribino2023role} and system responses to input with varying (im)politeness levels \cite{yin2024should}. Polite systems are generally perceived as more trustworthy, reliable, and acceptable \cite{ribino2023role}. LLMs have been shown to recognise (im)politeness, sometimes surpassing human-level performance \cite{li2023well, andersson2025can}, and to adapt their replies to the politeness level of user input \cite{quan2025human}. Politeness also influences system behaviour more subtly: across tasks and languages, polite prompts tend to elicit longer responses \cite{yin2024should}. 

User politeness in human–AI interaction has received comparatively less attention. Existing studies suggest that users differentiate between humans and systems in their use of indirectness, with the use of mitigation strategies reserved for human interlocutors \cite{lumer2023indirect}. Perceptions of system intelligence have also been found to shape user behaviour, with participants being more likely to describe themselves as polite toward more advanced AI systems \cite{yuan2024makes}. While some research reduces politeness to the presence of markers such as \textit{please} and \textit{thank you} \cite{barko2020conversational, yuan2024makes}, more detailed research into the impact of gender highlights systematic variation, with women tending towards hedging and men towards direct request strategies in task-oriented ChatGPT interactions \cite{tore2025speaking}.

Variation in conversational style, system politeness, and user politeness all shape human–AI interaction, but existing research remains fragmented. Some studies reduce politeness to politeness markers or subjective impressions, and others examine isolated prompts rather than extended information-seeking conversations. Our limited understanding of more comprehensive interactional styles and politeness patterns restrict our ability to effectively and inclusively design GenAI systems. Systems that respond only to particular politeness patterns may advantage some users while disadvantaging others. A systematic investigation of politeness strategies in conversational search settings is therefore needed—an agenda our work aims to advance.

%% file: 03_frummet_analyses.tex
\section{Analysing Frummet et al's data}\label{sec:aslett}

In this section, we analyse data from Frummet et al.~\cite{frummet2024cooking}, who investigated user–assistant interaction in a cooking context using a \textit{Wizard-of-Oz} experiment. In the study, participants followed step-by-step recipes via a chat interface and were randomly assigned to either an \textbf{active} or \textbf{passive} assistant condition. The active assistant proactively offered relevant background knowledge and assistance; the passive assistant responded only to direct user requests.

Participants were recruited via Prolific (native English speakers from the UK and US), and the recipe set consisted of six diverse and instructional meals. The original data, including all user utterances and system responses, were annotated and made publicly available.\footnote{\url{https://github.com/AlexFrummet/cooking-with-conversations}}

 In the following subsections we outline our politeness annotation process, then present cluster visualisations, and qualitative descriptions of the clusters.%, and tests for demographic effects.

\subsection{Annotation and Reliability}

In our analyses, we focus on the 30 conversations in the \textbf{passive}  condition, as this interaction style reflects how most current GenAI systems operate. In contrast, the active conversations more closely resembled human–human interaction, with the agent deliberately steering the exchange (e.g., prompting the user to consider specific steps in a cooking process, such that their curiosity for further information may be peeked).

We annotated utterances produced by both users and the agent in the passive condition. Each utterance was independently coded for the presence or absence of predefined politeness-related codes by a linguist. %A second annotator did the same for a sub-set of the data for validation purposes (see Section \ref{sec:validation}). 
These codes were developed based on Leech's \cite{leech2014pragmatics} classification scheme of politeness-sensitive speech acts, reflecting a goal-oriented approach to politeness, which is appropriate for our task-based cooking conversations with an assistant. Leech's scheme was slightly adapted and extended to capture all speech acts found in the interactions between the participants and the agent. As this study focusses on user politeness behaviour, the user utterances were annotated for the following categories:\footnote{Full details of the classification scheme can be found in the repository.}

\begin{itemize}
\item \textbf{Requests} are speech acts in which a speaker asks the addressee to perform a future action that entails a cost to them. Since requests restrict the addressee's freedom of action, speakers may soften them, e.g., by presenting them as optional, to enhance their politeness. According to Leech \cite{leech2014pragmatics}, requests comprise a spectrum of directive speech acts, stretching from direct commands to mild suggestions. Since people phrase requests with different levels of directness and in more or less conventional ways, we coded them using five distinct strategy types. \begin{enumerate}
\item \textbf{Direct requests} such as commands directly ask for the requested action, typically without mitigation (e.g., \textit{Show me how to chop the onions}).
\item \textbf{Non-sentential requests} also directly state the requested action but are conveyed without using a complete sentence (e.g., \textit{Next step}).
\item \textbf{Conventionally indirect requests} give the addressee the option not to comply (e.g., \textit{Can you show me...}).
\item \textbf{Statement hints} do not directly mention the requested action, which must be inferred (e.g., \textit{I'm ready to start}).
\item \textbf{Information-eliciting questions} \footnote{Information-eliciting questions have been added to Leech's scheme.} encompass questions aiming at eliciting information from the interlocutor without asking whether they are able or willing to do this. By signalling the speaker’s engagement and interest in continuing the interaction, they can increase politeness and help maintain harmony between interactants \cite{peters2005engagement, leech2014pragmatics} (e.g., \textit{When do I add the stock?; Do you have a video for this step?}). 
\end{enumerate}

\item \textbf{Politeness marker \textit{please}} captures the use of \textit{please} to mitigate the imposition of requests (e.g., \textit{Next step, please}).

\item \textbf{Apologies} intend to restore harmony between the speaker and addressee after an offence perceived as harmful to the speaker. In Frummet et al.'s data, there is only one apology by a participant, realised by \textit{sorry}.

\item \textbf{Absolving} codes the acknowledgement of apologies by the user by (a) \textbf{accepting} them (e.g., \textit{That's ok}), (b) \textbf{denying} the reality of the offence (e.g., \textit{No problem}), or (c) \textbf{ignoring} the offence due to its triviality (e.g.,  \textit{Don't mention it}).

\item \textbf{Thanking} is a face-enhancing expression of gratitude, for instance, for helpful information (e.g., \textit{Thank you for the info}).
    
\item \textbf{Compliment/praise/face-enhancing feedback} is an umbrella category comprising face-enhancing speech acts that positively evaluate the agent or its contributions (e.g., \textit{Excellent}; \textit{That’s a great tip}; \textit{This is really useful}).
    
\item \textbf{Acknowledgements of responses} \footnote{Acknowledgments of responses have been added to Leech's scheme.} include user reactions signalling receipt and understanding of the agent's response, thus showing their engagement (e.g., \textit{I see; ok}). Acknowledgments including a positive evaluative comment (e.g., \textit{That’s helpful}) are additionally coded as face-enhancing feedback.

\item \textbf{Declining} codes speech acts through which users turn down offers by the agent (e.g., \textit{No thank you}).

\end{itemize}

In total 14 unique codes were applied to user utterances. Each utterance could be assigned multiple codes if more than one pragmatic function was present.

\subsection{Validating the Annotation Process}
\label{sec:validation}
To evaluate inter-rater reliability, a second linguist recoded a random sample of 20\% of the data (6 conversations). We computed \textbf{Krippendorff’s alpha} for each user code. %, treating \textbf{user} and \textbf{system} utterances separately.
  Each utterance was coded in a binary fashion, either as an \textit{application} of the code (present) or a \textit{non-application} (absent), and reliability was calculated across all utterances. %Values ranged from $ \alpha = .39 $ to $ \alpha = 1.00 $, with 15 out of 20 codes achieving $ \alpha > .65 $. For subsequent analyses, we focused on the 14 \textbf{user-directed} politeness codes, which on average showed higher reliability than the \textbf{agent-directed} codes. 
  The mean $ \alpha = .769$. All but one code $ \alpha > .60 $. These included \textit{Thanking}, \textit{Politeness marker please}, and \textit{Compliment/praise}, which all reached perfect agreement ($ \alpha = 1.00 $). The only code  $ \alpha < .60 $ was \textit{Request -- direct}, which showed lower agreement ($ \alpha = .39 $). This was due to the low overall frequency of direct requests (N=6) and a perceived overlap with non-sentential requests. One coder annotated \textit{start} as \textit{Request - direct} and the other as \textit{Request - non-sentential}, an easy issue to resolve.

\subsection{User Representation and Clustering}
For each user, we computed a vector of the absolute frequencies of the selected politeness codes, aggregated across the annotated conversations. These frequencies were standardised (z-scores), producing a matrix, where each row represented a user and each column a code. We applied \textbf{k-means clustering} to group users based on their code usage profiles. The optimal number of clusters (\(k\)) was determined using the \textbf{elbow method}, which suggested \(k = 3\) or \(4\).
We opted for \(k = 4\) to better capture nuanced behavioural patterns, despite one cluster containing a single user, because this cluster represents a theoretically meaningful extreme of politeness behaviour that is rare in our small sample but plausibly more prevalent in the wider population.

\subsection{Visualisation and Interpretation}
%We used \textbf{t-distributed Stochastic Neighbour Embedding (t-SNE)} to reduce the high-dimensional code space to two dimensions. Each user was projected into this 2D space and colour-coded by cluster membership. Labels were anonymised (e.g., U01--U30) for interpretability.

To interpret cluster characteristics, we calculated the \textbf{mean standardised usage} of each politeness code per cluster, visualising these means using bar plots to highlight which speech acts were more or less prominent in each cluster (Figure \ref{fig:cluster_vis}). This informed our characterisation of clusters based on the politeness and interactional strategies employed. %Based on clustering, we identified four distinct user clusters relating to the politeness and interactional strategies employed. 
Below, we summarise the main characteristics of each cluster providing examples from the data:

% as described in the upcoming subsection.

\begin{figure}[htbp]
    \centering
    \includegraphics[width=\linewidth]{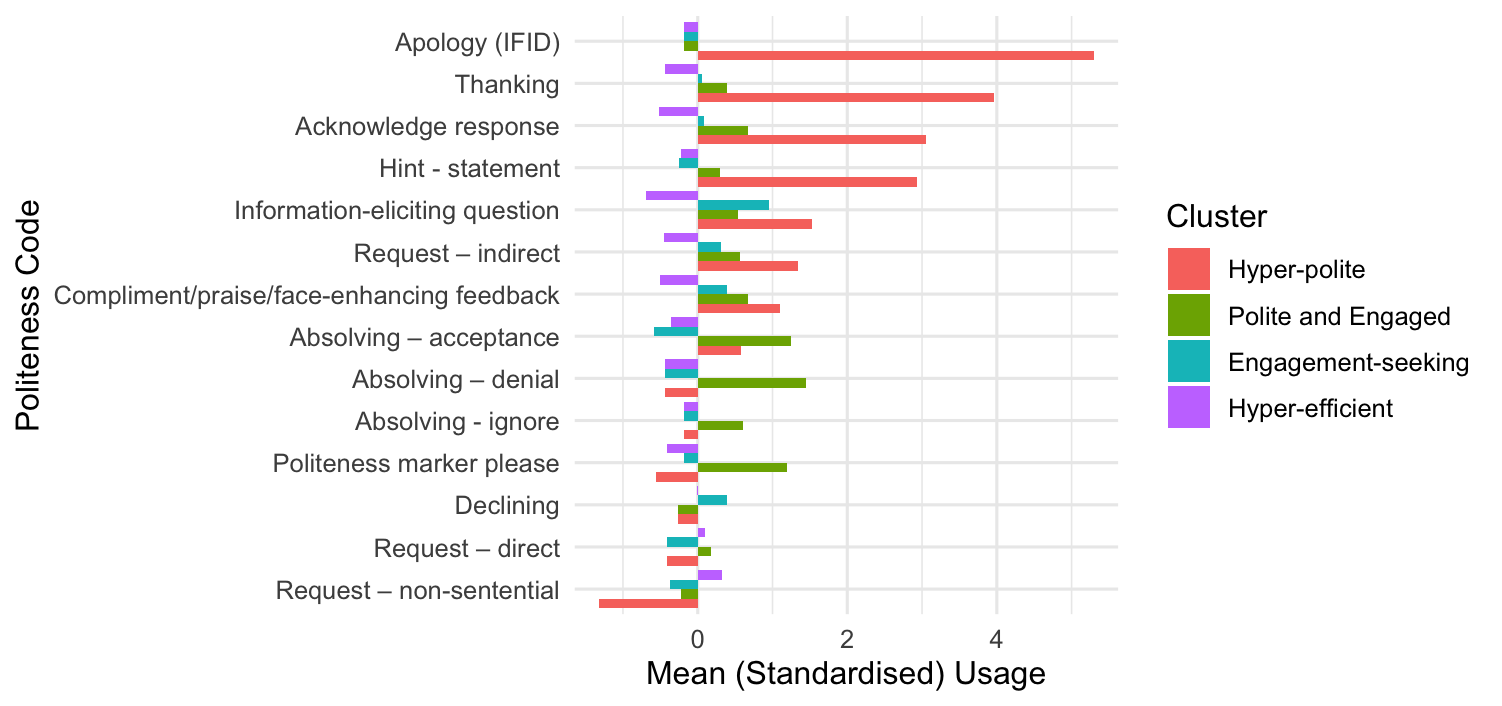}
    \caption{Z-score standardised code frequency across clusters}
    \label{fig:cluster_vis}
\end{figure}
%\subsection{Summary of Politeness Clusters}

%Based on clustering, we identified four distinct user clusters relating to the politeness and interactional strategies employed. Below, we summarise the main characteristics of each cluster providing examples from the data:

\subsubsection*{Hyper-polite cluster}
This cluster is characterised by a very politeness-sensitive and engaged interactional style
but it only comprises one participant. Thanking, face-enhancing feedback, acknowledging responses (e.g., \textit{that is interesting}; \textit{looks good}; \textit{okay}), and information-eliciting questions (e.g., \textit{do the basil leaves need to be organic?}) are frequently used. Less frequent but notable are indirect requests (e.g., \textit{Can you show me what it would look like?}) and accepting agent apologies (e.g., \textit{ok}). The participants avoids face-threatening direct and non-sentential requests, opting for more indirect statement hints to move along in the recipe instead (e.g., \textit{I’m ready for the next step}) but does not use the politeness marker \textit{please} at all. Notably, this is the only participant who apologises to the agent, in a repair of a preceding information-eliciting question (\textit{what is the quantity of it sorry?}). These patterns suggest a very polite and engaged communication style catering to the addressee's positive face needs by showing an interest in the information they provide and offering face-enhancing feedback. This style is indicative of treating the agent more like a human interlocutor than a machine.

\subsubsection*{Polite and engaged cluster}
This cluster is marked by a polite, face-enhancing, and engaged communication style that is employed with moderate frequency. It contains seven participants who fairly regularly use both face-threat mitigating strategies, such as indirect requests (e.g., \textit{can you show me a picture}) and statement hints (e.g., \textit{I think i'm ready to move to the next step}), and face-enhancing strategies, including acknowledging responses, thanking and compliments/face-enhancing feedback (e.g., \textit{Awesome, thank you}), as well as absolving reactions to agent apologies. Notably, they use absolving strategies more often and with greater variation than other clusters (accepting (e.g., \textit{that's okay}), ignoring the offence (e.g., \textit{Doesn't matter}), denying the reality of the offence (e.g., \textit{no worries})). The politeness marker \textit{please} is also more common than in any other cluster, often paired with indirect requests to further mitigate imposition (e.g., \textit{can you tell me about the history of this recipe please?}) and with non-sentential requests to proceed  (e.g., \textit{Thanks, next step please}). Overall, the participants exhibit a polite, respectful and appreciative communication style, showing an interest in additional information.

\subsubsection*{Engagement-seeking cluster}
The interactional style in this cluster is distinguished by a balance of politeness and efficiency. It includes six participants who frequently engage with the agent, particularly through varied information-eliciting questions (e.g. \textit{where is the note about pecorino fiore sardo}; \textit{Do I need to remove the zest?}), alongside a measured use of thanking, acknowledging responses, indirect requests, and face-enhancing feedback. These users tend to avoid statement hints, absolving responses to apologies, and direct requests, while using non-sentential requests to a limited degree. These rarely combine with the politeness marker \textit{please}, which is not used in any other context. Their communication style reflects a preference for smooth, efficient interactions with high engagement and the occasional use of face-enhancing strategies.

\subsubsection*{Hyper-efficient cluster}
This cluster exhibits low engagement and minimal use of politeness-related strategies such as thanking, acknowledging responses, face-enhancing feedback, or absolving reactions to apologies. It comprises 16 participants who use the fewest information-eliciting questions among all clusters and frequently rely on non-sentential requests (e.g., \textit{next}) to move the process along, usually without mitigating devices like the politeness marker \textit{please}. While they occasionally use indirect requests, they also sometimes resort to direct requests (e.g., \textit{Remind me which types of cheese to use}). Their communication style is direct, task-oriented, and efficiency-driven, reflecting a tendency to treat the agent as a machine rather than a human conversational partner.\\

The politeness patterns manifested in the clusters represent a continuum from hyper-polite to hyper-efficient with two intermediate profiles reflecting distinct interactional goals and behaviour.

\begin{comment}
\subsection{Demographic Patterns in Politeness}

For the demographic analysis, we merged the conversation metadata with cluster assignments. Since cluster 1 contained only a single participant, it was excluded from further analysis, as meaningful patterns cannot be inferred from a group of one. The remaining clusters were then compared by age and gender. To address small cell sizes in the original age categories, we collapsed them into three broader groups: 18–34, 35–54, and 55+.

Across the three clusters, the majority of participants fell into the 18–54 age range, with very few participants aged 55 or older. Gender distributions were similar across clusters, with a slight overrepresentation of women in the \textit{Engagement-seeking} and \textit{Polite-engaged} clusters and a more balanced distribution in the \textit{Hyper-efficient} cluster. These results suggest broadly comparable demographic profiles across clusters. Unfortunately, it was not possible to link geographic location or income information to the conversations in Frummet et al.’s dataset.
\end{comment}

%% file: 04_simulation.tex
\section{Simulating (Im)polite Conversations}\label{sec:simulation}

To learn whether 
different (im)politeness strategies shape interactions with generative AI systems we performed a pre-registered experiment\footnote{The pre-registration can be found here: \url{https://osf.io/mp3dt/overview}}%\url{https://osf.io/mp3dt/?view_only=40180eb3f2634239b8d4485e3eb094d8}} 
using a controlled simulation in which two LLMs interacted: 
one acting as a \emph{user} and the other as the \emph{agent}. The following sub-sections outline the precise method employed, the hypotheses tested, the analysis plan, and the results.

\subsection{Simulation Approach}

%It has been shown that the Principle of (Im)politeness Reciprocity (PIR) applies to both face-to-face and online interactions \cite{culpeper2021principle,culpeper2025impoliteness}, i.e., people tend to reciprocate the level of (im)politeness they encounter. As models such as ChatGPT have been trained to “avoid offensive content” \cite{andersson2025can, dynel2023lessons}, 

\paragraph{Recipe task domain.} 
We adopted the same six recipes used by Frummet et al. \cite{frummet2024cooking}, all sourced from SeriousEats 
(\url{https://www.seriouseats.com}): 
\emph{Parisian Gnocchi}, \emph{Buttermilk-Brined Southern Fried Chicken}, \emph{Duck à l’Orange}, 
\emph{Savory Cheese Soufflé}, \emph{Pesto alla Genovese}, and \emph{Old-Fashioned Apple Pie}. 
Each recipe was decomposed into stepwise instructions (e.g., \textit{score the duck skin and season with salt}). 
For each step we generated a set of neutrally phrased information needs spanning three categories: 
(1) task-related (e.g., \textit{How deep should the scoring be?}), 
(2) science-related (e.g., \textit{How does scoring influence fat rendering?}), and 
(3) history-related (e.g., \textit{When did béchamel originate?}). 
These needs served as content ``slots'' that could be realised in different pragmatic styles.

\paragraph{User profiles.} 
To operationalise variation in politeness, we defined multiple user profiles derived from our corpus analysis (i.e., \emph{Hyperpolite}, \emph{Polite and engaged}, \emph{Engagement-seeking}, and \emph{Hyper-efficient}). We additionally constructed a fifth profile, 
\emph{Impolite}. 
Although we did not observe impolite behaviour in the naturalistic conversations collected by Frummet et al., there is evidence of users interacting impolitely with GenAI systems \cite{tore2025speaking}. We added this profile to test how agents would react to impolite interactions in this context. 
The impoliteness features for this profile are based on a selective combination of Culpeper's impoliteness output strategies \cite{culpeper1996towards} and conventional impoliteness formulae \cite{culpeper2011impoliteness, culpeper2017impoliteness} that are relevant to the task-based cooking interaction between a user and an agent. The \emph{Impolite} cluster is marked by face-threatening acts in the shape of imperatives and brusque commands and instances of withholding politeness through a lack of thanking, acknowledgments, or compliments. It moreover includes criticisms and disagreements (e.g., \textit{This is irrelevant!}) as well as insults and taboo words (e.g., \textit{You idiot}, \textit{That is total crap!}) and is characterised by
 frequent abrupt moves to the next step 
without engagement.
Each profile was characterised by probabilistic tendencies to use particular politeness strategies. 
These tendencies were described to the user-LLM in its system prompt, along with instructions to output turns 
in valid JSON containing: the utterance, the politeness codes expressed, the pragmatic intent, 
and whether the user wished to proceed to the next recipe step. 
This allowed us both to guide the LLM toward a target style and to log the realised behaviour.

\paragraph{Agent behaviour.} 
The agent-LLM received the recipe step, the user’s most recent utterance, and the conversation history. 
Its prompt instructed it to act as a  knowledgeable cooking assistant, answering task, science, 
or history questions as they arose. 
In approximately fifteen percent of cases, particularly for science or history needs, the agent was prompted to 
\emph{fail gracefully} by apologising for not having an answer. 
This enabled us to simulate how users in different clusters responded to agent apologies. 
Like the user, the agent’s output was constrained to JSON, containing the text of its reply and a list of 
politeness codes expressed in the turn.

\paragraph{Conversation flow.} 
Each recipe step was simulated as a short dialogue loop. 
The user-LLM produced an utterance guided by its politeness profile and the selected information need. 
The agent-LLM then responded with an answer, or, in some cases, with an apology. 
The loop continued until either the user signalled that they wished to proceed or a maximum of 
two questions had been asked about that step. This mirrors the conversations in Frummet et al.'s data.
All turns were stored with their associated metadata, including a chronological event index.

\paragraph{Data collection.} 
The simulation produced structured dialogue logs in tabular form. 
Each row corresponds to one utterance, annotated with: recipe step, need type, role (user vs.\ agent), 
text, pragmatic intent, politeness codes, and, where relevant, apologies. 
These logs allow us to (a) validate whether the generated user behaviours cluster according to the 
intended politeness profiles and
(b) quantify the amount of information transferred and energy used under different conditions. %, and 
%(c) analyse how users react to agent apologises for not being able to answer.

\paragraph{Models.} 

 We selected three open-weight instruction-tuned models in the 7--8B parameter class: \textit{DeepSeek-8B}, \textit{Llama-3.1-8B-Instruct}, and \textit{Qwen-2.5-7B-Instruct}. This size range ensures comparability in computational demands while avoiding scale effects. The models represent distinct training lineages and design aims: DeepSeek is a distilled model optimised for reasoning and instruction following \cite{deepseek2024}, Llama-3.1 reflects Meta's broadly aligned and stylistically adaptive approach to multilingual assistant chat \cite{llama2024}, and Qwen-2.5 incorporates extensive multilingual and cross-cultural training data with supervised fine-tuning and RLHF \cite{qwen2024}. This diversity makes them well suited for investigating how user politeness strategies influence model responses.

Full code, including user- and agent-prompts, as well as the conversations themselves can be found in our repository. 

\paragraph{Validation.} To validate the approach, a set of 300 test conversations were simulated and 50 sampled to cover all 5 politeness profiles. Two linguists independently labelled the conversations based on the politeness profiles (clusters) employed by the user bot. The annotators achieved an accuracy 92\% and 96\%, respectively, which we accept as evidence for the simulations of the clusters behaving authentically.

\paragraph{Sample size} 
For the final experiment, we generated 18,000 conversations (5 politeness profiles x 6 recipes x 3 models x 200 conversations per cell). To determine this number we conducted Monte Carlo power simulations using conversation-level negative binomial models. %(nuggets\_sum $\thicksim$ cluster × model + recipe + offset(log Wh\_sum)).
Parameters calibrated from the 300 test conversations generated step-level data with a conversation random effect, then we aggregated to conversation-level for analysis. Power was the proportion of simulated datasets with a significant cluster × model term ($\alpha = .05$). Power simulations indicated that 200 conversations per cluster--recipe--model cell achieved 83\% power (95\% CI [77\%, 90\%]) to detect the planned interaction ($\sim\!\pm 2\%$). %An MDE sweep showed that with 150--200 per cell the minimum detectable interaction is approximately $\pm 3\%$ ($\geq 80\%$ power).

\subsection{Hypotheses}

Based on prior work in pragmatics and conversational style, as well as our initial simulations, 
we expect systematic differences across politeness profiles in how conversations unfold. 
Each hypothesis operationalises one research question (RQ1--RQ3).

\paragraph{H1: Response length.} 
The length of agent responses (in tokens) will vary across politeness profiles. 
Polite or hyperpolite inputs may encourage more elaborated responses, whereas efficient and direct or impolite inputs may 
yield shorter, more task-focused answers.

\paragraph{H2: Information transfer.} 
The number of new ``information nuggets'' per recipe step---defined as atomic, useful pieces of information not 
previously mentioned in the same utterance---will vary across politeness profiles.

\paragraph{H3: Energy efficiency.} 
The efficiency of information transfer, operationalised as the ratio of information nuggets to energy consumption (Wh), 
will vary across politeness profiles.

\subsection{Analysis Plan}

The presented analyses follow the pre-registered analysis plan. Below we outline the dependent variables and how they are measured:

\paragraph{Response length.} 
We measure the token length of agent responses in words.

\paragraph{Information transfer.} 
To measure the amount of information conveyed, we adopted a nugget-counting approach similar to Frummet et al., where an LLM is used to identify and count minimal factual units (“nuggets”). 
In our implementation, agent utterances were concatenated per step, which required handling duplicate information. 
We employed a few-shot prompting setup with the \texttt{deepseek} model to generate nugget counts.  

We validated the nugget counter against a manually annotated test set ($N = 158$ steps) with two coders, covering edge cases such as no information (e.g., apologies), duplicated content (rephrasings), mixed procedural and declarative knowledge, and high variability in information density. Model outputs correlated well with human judgments ($r = .70$, MAE $= 0.80$, bias $= -0.22$) and showed excellent consistency across three runs ($\mathrm{ICC}(C,3) = 0.994$, 95\% CI [0.992, 0.995]); for comparison, inter-coder agreement was $r = .73$. This indicates that the metric is both valid and reliable, providing a stable basis to compare information density across conditions.

\textit{Energy efficiency.} 
We compute an efficiency score (nuggets per Wh consumed).% by regularly sampling the graphics card for its energy use.

\section{Results}

\begin{table*}[t]
\centering
\caption{Back-transformed expected \emph{words per step} (H1) and \emph{nuggets per step} (H2), plus \emph{nuggets per 100 words} (H3a) and \emph{nuggets per Wh} (H3b), by Cluster $\times$ Agent. 95\% CIs shown as [LCL:UCL]. Estimates adjust for recipe and conversation clustering.}
\label{tab:merged_words_nuggets_density_energy}
\small
\setlength{\tabcolsep}{5pt}
\begin{tabular}{llllll}
\toprule
Agent & Cluster & Words [LCL:UCL] & Nuggets [LCL:UCL] & Nuggets/100w [LCL:UCL] & Nuggets/Wh [LCL:UCL] \\
\midrule
deepseek & Hyperpolite & 69.9 [69.1:70.7] & 2.03 [2.00:2.06] & 1.61 [1.59:1.64] & 1.654 [1.629:1.680] \\
deepseek & Polite-Engaged & 50.2 [49.5:50.9] & 1.87 [1.84:1.91] & 1.92 [1.88:1.95] & 1.979 [1.941:2.017] \\
deepseek & Engagement-Seeking & 76.9 [76.0:77.8] & 2.16 [2.13:2.19] & 1.61 [1.58:1.63] & 1.598 [1.574:1.622] \\
deepseek & Hyperefficient & 42.6 [42.1:43.1] & 1.59 [1.56:1.62] & 1.93 [1.90:1.96] & 1.858 [1.826:1.891] \\
deepseek & Impolite & 66.2 [65.5:67.0] & 1.73 [1.70:1.76] & 1.40 [1.37:1.42] & 1.331 [1.309:1.353] \\
\midrule
llama & Hyperpolite & 67.9 [67.1:68.7] & 2.11 [2.07:2.14] & 1.79 [1.76:1.81] & 2.040 [2.010:2.071] \\
llama & Polite-Engaged & 53.2 [52.5:54.0] & 2.03 [1.99:2.07] & 2.15 [2.11:2.19] & 2.527 [2.481:2.574] \\
llama & Engagement-Seeking & 69.8 [69.0:70.6] & 2.26 [2.23:2.29] & 1.91 [1.88:1.94] & 2.075 [2.045:2.105] \\
llama & Hyperefficient & 42.8 [42.3:43.3] & 1.87 [1.84:1.90] & 2.51 [2.47:2.55] & 2.702 [2.659:2.746] \\
llama & Impolite & 63.2 [62.4:63.9] & 2.11 [2.08:2.14] & 1.94 [1.91:1.97] & 2.005 [1.975:2.035] \\
\midrule
qwen & Hyperpolite & 53.0 [52.4:53.6] & 1.93 [1.90:1.96] & 2.08 [2.05:2.11] & 1.968 [1.937:1.999] \\
qwen & Polite-Engaged & 38.3 [37.7:38.8] & 1.68 [1.65:1.72] & 2.39 [2.34:2.44] & 2.251 [2.206:2.297] \\
qwen & Engagement-Seeking & 57.5 [56.8:58.1] & 2.02 [1.98:2.05] & 2.03 [2.00:2.06] & 1.907 [1.878:1.936] \\
qwen & Hyperefficient & 30.3 [30.0:30.7] & 1.46 [1.43:1.49] & 2.45 [2.40:2.49] & 2.144 [2.105:2.183] \\
qwen & Impolite & 44.4 [43.9:44.9] & 1.63 [1.61:1.66] & 1.98 [1.95:2.02] & 1.642 [1.614:1.670] \\
\bottomrule
\end{tabular}
\end{table*}

Following the pre-registered analysis plan, we analyse step-level outcomes with mixed-effects models that include \emph{Cluster} (user politeness profile) and \emph{AgentModel} as fixed factors (controlling for \emph{Recipe}) and a random intercept for \emph{Step} to account for the repeated measure (steps within a conversation). For H1–H3, we first run omnibus tests (e.g., ANOVA or analogue on the fitted mixed models) to determine whether \emph{Cluster}, \emph{AgentModel}, or their interaction explains outcome variance. If this is significant we conduct follow up pair-wise comparisons of estimated marginal means with Tukey adjustment. % for multiple comparisons.
For accessibility, each hypothesis subsection is structured as follows: 1. Non-technical summary of the main take-away message, 2. Detailed statistical methods, 3. Detailed findings and commentary.\footnote{For space reasons only key details are reported here. Full models and descriptives are available in our repository. \url{https://github.com/delsweil/CHIIR2026_politeness}}

\subsection{H1: Do user politeness profiles change how long agents reply?}
\label{sec:H1}

\textbf{Take-away.} Politeness profiles systematically change reply length: \emph{Engagement-seeking (ES)} prompts elicit the longest answers; \emph{Hyperefficient (HE)} the shortest; \emph{Polite and engaged (PE)} and \emph{Hyperpolite (HP)} sit in between, with \emph{Impolite (IMP)} typically shorter than PE.\\

\noindent\textbf{What we tested \& why.}
We modelled reply length at the step level using a linear mixed-effects model on $\log(\text{words}+1)$ with \textit{Cluster}, \textit{AgentModel}, and \textit{Recipe} as fixed effects and a random intercept for \textit{Step} (steps within a conversation are not independent). \textit{Recipe} controls for content differences. Lengths are right-skewed; the log transform reduces the influence of extremes and makes effects interpretable as proportional (percentage) differences. For interpretability we report back-transformed estimated marginal means (EMMs) as \emph{words per step}.\\

\noindent\textbf{Omnibus tests.}
Mixed-effects ANOVA found reliable effects:
Cluster: F$_{4,16029}=4450.29$, $p<.001$;
AgentModel: F$_{2,16668}=3981.49$, $p<.001$;
Cluster$\times$AgentModel: F$_{8,16026}=47.16$, $p<.001$.\\

%\noindent\textbf{Model-based means (what changes, practically).}
Table~\ref{tab:merged_words_nuggets_density_energy} shows back-transformed EMMs for words (H1). Qwen produced on average shorter responses than the other models. However, across agent models consistent trends are apparent: ES has the longest replies and HE the shortest, with large within-model differences (see \autoref{tab:quick_all} for illustrative examples of how deltas vary across models and profiles).

% --------------------------------------------------------------

\subsection{H2: Do politeness profiles change how many nuggets agents convey?}
\label{sec:H2}

\noindent\textbf{Take-away:} ES prompts yield the most nuggets per step, HE the fewest; PE generally exceeds IMP. However, \emph{percentage increases in words are much larger than percentage increases in nuggets.} Some agents appear less open to influence from impoliteness on \emph{information quantity}, even when their \emph{reply length} still shifts.\\

\noindent\textbf{What we tested \& why.}
Nuggets are non-negative counts and were overdispersed, so we used a negative binomial mixed model (NB--GLMM) with \textit{Cluster} and \textit{AgentModel} as fixed effects, \textit{Recipe} as a control, and a random intercept for \textit{Conversation}.\\

\noindent\textbf{Omnibus tests.}
Omnibus effects were tested via likelihood--ratio $\chi^2$ tests comparing nested NB--GLMMs ($\Delta\text{deviance}=2\,\Delta\log\text{-likelihood}$; df equals the number of fixed-effect parameters removed).
Results indicated strong main effects of \textit{Cluster} and \textit{AgentModel}, and a reliable \textit{Cluster}$\times$\textit{AgentModel} interaction:
Cluster: $\chi^2_{(4)}=1895.38$, $p<.001$; AgentModel: $\chi^2_{(2)}=1087.78$, $p<.001$; Cluster$\times$AgentModel: $\chi^2_{(8)}=239.22$, $p<.001$.\\

%\noindent\textbf{Model-based means (what changes, practically).}
Table~\ref{tab:merged_words_nuggets_density_energy} reports EMMs for \emph{nuggets per step} on the original count scale. The pattern is consistent across agent models: ES produces the most nuggets, HE the least, and PE exceeds IMP. Magnitudes vary by agent model, but the direction is stable. The overview in \autoref{tab:quick_all} quantifies these gaps, which can be up to 80.5\% in words and 38.2\% in nuggets.\\

\noindent\textbf{Pairwise exceptions (after Tukey).}
Most cluster comparisons were significant after Tukey adjustment. The non-significant pairs were: llama: HP vs.\ IMP; qwen: PE vs.\ IMP. This may suggest that, for these models, \emph{impoliteness did not reliably reduce the amount of information (nuggets per step)} relative to those polite styles. We interpret this as \emph{reduced sensitivity to impoliteness in information quantity}, with three caveats: (i) this is \emph{outcome–specific} (H1 shows length still changes, and H3 addresses efficiency); (ii) it is \emph{pair–specific} (other IMP comparisons were significant); and (iii) non–significant $\neq$ equivalence.\\%—CIs are compatible with small positive or negative effects that our multiplicity–controlled tests do not resolve.\\

\noindent\textbf{Words vs.\ nuggets.} Our analyses in this subsection highlight that, across models, moving from HE to ES yields \emph{much larger percentage increases in words} than in nuggets: roughly 60--90\% more words for only about 20--40\% more nuggets (see \autoref{tab:quick_all}). Likewise, IMP often elicits \emph{longer} replies than PE with only minimal nugget differences (approximately $-4\%$ to $+8\%$). In short, \emph{length grows faster than information}. This relates to H3, where we analyse \emph{information density and energy efficiency} (nuggets per word; nuggets per Wh) in detail.

% Requires \usepackage{booktabs, tabularx, makecell}
\begin{table}[t]
\centering
\caption{Within-agent gaps across outcomes for two illustrative pairwise comparisons. Entries show $\Delta$ and \% (relative to the second cluster). H1 = words; H2 = nuggets; H3a = nuggets/100w; H3b = nuggets/Wh.}
\label{tab:quick_all}
\begingroup
\scriptsize
\setlength{\tabcolsep}{3pt}
\renewcommand\arraystretch{1.05}
\begin{tabularx}{\columnwidth}{l
  >{\raggedright\arraybackslash}p{0.38\columnwidth}
  >{\raggedright\arraybackslash}p{0.38\columnwidth}}
\toprule
Agent & ES--HE (W; N; N/100w; N/Wh) & PE--IMP (W; N; N/100w; N/Wh)\\
\midrule
deepseek &
\makecell[tl]{W: 34.3 (80.5)\\
             N: 0.57 (35.9)\\
             N/100w: $-32.3$ ($-16.7$)\\
             N/Wh: $-0.26$ ($-14.0$)} &
\makecell[tl]{W: $-16.1$ ($-24.2$)\\
             N: 0.14 (8.2)\\
             N/100w: 51.9 (37.1)\\
             N/Wh: 0.65 (48.7)}\\
\addlinespace[2pt]
llama &
\makecell[tl]{W: 27.0 (63.2)\\
             N: 0.39 (20.7)\\
             N/100w: $-59.9$ ($-23.8$)\\
             N/Wh: $-0.63$ ($-23.2$)} &
\makecell[tl]{W: $-9.9$ ($-15.7$)\\
             N: $-0.08$ ($-3.6$)\\
             N/100w: 20.6 (10.6)\\
             N/Wh: 0.52 (26.0)}\\
\addlinespace[2pt]
qwen &
\makecell[tl]{W: 27.1 (89.4)\\
             N: 0.56 (38.2)\\
             N/100w: $-41.8$ ($-17.1$)\\
             N/Wh: $-0.24$ ($-11.1$)} &
\makecell[tl]{W: $-6.2$ ($-13.9$)\\
             N: 0.05 (3.1)\\
             N/100w: 40.3 (20.3)\\
             N/Wh: 0.61 (37.1)}\\
\bottomrule
\end{tabularx}
\endgroup
\end{table}

\subsection{H3a: Do politeness profiles change information density?}
\label{sec:H3a}

\noindent\textbf{Takeaway.} Hyperefficient (HE) prompts yield the \emph{densest} replies (more nuggets per 100 words), Engagement-seeking (ES) the \emph{least dense}; Polite and engaged (PE) is consistently denser than Impolite (IMP).

\noindent\textbf{What we tested \& why.}
We analysed step‐level \emph{nugget counts} with a negative–binomial mixed model (log link). We included log(words) as an \emph{offset}, which means effects are estimated on a \emph{rate} scale (how many nuggets are produced \emph{per word}) rather than on raw counts. Fixed effects were the \emph{Cluster × AgentModel} interaction and \emph{Recipe} (as a control), and we added a random intercept for \emph{Step} to account for repeated steps within conversations. Exponentiating coefficients gives \emph{incidence–rate ratios}: values 
>1 indicate more nuggets per word than the reference level; values <1 indicate fewer.\\

%\noindent\textbf{Model-based means (practical differences).}
Table~\ref{tab:merged_words_nuggets_density_energy} shows nuggets per 100 words by Cluster $\times$ Agent. Across agents, HE has the highest density, while ES has the lowest.
Within-agent deltas (Table~\ref{tab:quick_all}) quantify the gaps. 
In short, ES tends to be \emph{word-inefficient}, while HE/PE concentrate information better.\\

\noindent\textbf{Omnibus Tests.} Likelihood-ratio $\chi^2$ tests show significant effects for Cluster: $\chi^2_{(4)}=2097.36$, $p$ < .001; AgentModel: $\chi^2_{(2)}=2633.61$, $p$ < .001; Cluster$\times$AgentModel: $\chi^2_{(8)}=335.62$, $p$ < .001.\\ 

\noindent\textbf{Pairwise exceptions (Tukey-adjusted).}
All but 6 density comparisons were significant. Non-significant pairs were:
deepseek: HE vs.\ PE; HP vs.\ ES.
llama: ES vs.\ IMP.
qwen: ES vs.\ IMP; HE vs.\ PE; HP vs.\ ES.
Unlike previous tests, there is no clear pattern here. Again we note the non-significant \(\neq\) equivalence.\\

\begin{figure*}[t]
  \centering
  \includegraphics[width=0.75\textwidth]{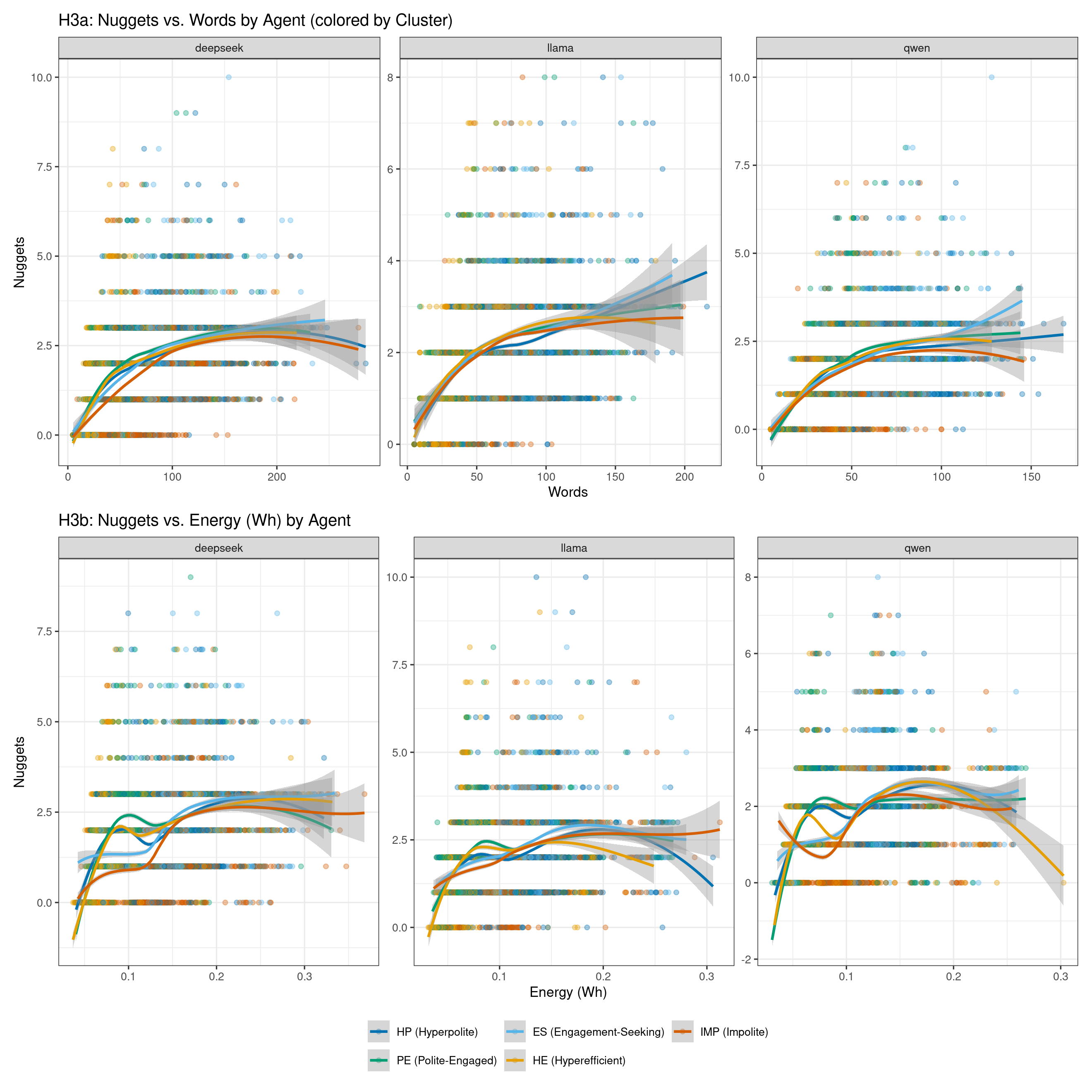}
  \caption{H3a/H3b efficiency diagnostics. Top: nuggets vs.\ words; bottom: nuggets vs.\ energy (Wh). Columns are agents; colours denote user clusters. LOESS smooths with 95\% CIs over step-level points. X-axes are free per column to reflect very different scales. Patterns show early gains followed by plateaus (diminishing returns).}
  \Description{Two-row, three-column faceted scatterplots with LOESS smooths, coloured by user cluster, demonstrating saturation of nuggets with additional words or energy across all agents.}
  \label{fig:h3_combined}
\end{figure*}

\noindent\emph{Note:} Our models fit means, however the LOESS curves in Fig.~\ref{fig:h3_combined} show that nuggets rise steeply over the first tens of words and then flatten. The plateau occurs at slightly different points for different models, but after 50-100 words, additional text yields progressively smaller gains (often less than half a nugget), indicating diminishing returns in informational yield per extra word. Clusters differ mainly by the amount of information they provide at a given length, not by the presence of a different slope.

\begin{comment}

\begin{figure*}[t]
  \centering
  \begin{subfigure}[t]{0.49\textwidth}
    \centering
    \includegraphics[width=\linewidth]{plots/scatter_nuggets_vs_words.png}
    \caption{H3a: Nuggets vs.\ words by Agent (coloured by Cluster). LOESS smooths with 95\% CIs.}
    \label{fig:h3a_scatter}
  \end{subfigure}\hfill
  \begin{subfigure}[t]{0.49\textwidth}
    \centering
    \includegraphics[width=\linewidth]{plots/scatter_nuggets_vs_energy.png}
    \caption{H3b: Nuggets vs.\ energy (Wh) by Agent (coloured by Cluster). LOESS smooths with 95\% CIs.}
    \label{fig:h3b_scatter}
  \end{subfigure}
  \caption{Information growth shows \emph{diminishing returns}. Across agents, nuggets rise quickly with initial words/energy, then flatten—extra verbosity or compute adds little new information beyond a point.}
  \label{fig:h3_scatter_both}
\end{figure*}
\end{comment}

\begin{comment}
\begin{figure}[t]
  \centering
  % swap .pdf for .png if that's what you saved
  \includegraphics[width=\columnwidth]{plots/H3_scatter_combined.png}
  \caption{H3a/H3b efficiency diagnostics. Top row: nuggets vs.\ words; bottom row: nuggets vs.\ energy (Wh). Columns are agents; colours denote user clusters. Curves are LOESS smooths with 95\% CIs over step-level points. X-axes are free per column (note very different scales for words vs.\ Wh). The curves rise quickly then plateau, indicating diminishing returns in nuggets with additional words/energy.}
  \Description{Two-row, three-column faceted scatterplots with LOESS smooths. Top row (words) shows rapid initial nugget gains that flatten across agents; bottom row (Wh) shows a similar saturation pattern. Colours encode user clusters.}
  \label{fig:h3_combined}
\end{figure}
\subsection{H3b: Do politeness profiles change energy efficiency?}
\label{sec:H3b}
\end{comment}

\subsection{H3b: Do politeness profiles change
energy efficiency?}
\noindent\textbf{Take-away.} IMP is the least energy-efficient across agents. PE reliably beats IMP; HE and PE are the most efficient; ES is typically less energy-efficient than HE.\\

\noindent\textbf{What we tested \& why.}
We modelled the \emph{nugget count} per step with a negative–binomial mixed model (log link) and included log(Wh) as an \emph{offset}. The offset treats watt-hours as the exposure, so all effects are estimated on a \emph{rate} scale (i.e., how many nuggets are produced \emph{per unit of energy}). In other words, the model asks: holding energy usage constant, do different user clusters and agent models yield different \emph{nuggets/Wh}? We used negative binomial because nugget counts are overdispersed (variance exceeds the mean). Fixed effects were the \emph{Cluster × AgentModel} interaction (our focal factor) and (as in previous models) \emph{Recipe} as a control for content differences. We added a random intercept for \emph{Step}.\\

%\noindent\textbf{How to read the estimates.}
%Exponentiated coefficients from this model are \emph{incidence-rate ratios} (IRRs) on the \emph{nuggets/Wh} scale (values >1 mean more nuggets per watt-hour than the reference; <1 mean fewer). For interpretability, we report \emph{estimated marginal means} (EMMs) on the response scale as \emph{nuggets per Wh} with 95\% confidence intervals, which are the model-based expected rates after averaging over recipes and accounting for conversation clustering.

%\noindent\textbf{Model-based means (practical differences).}
Table~\ref{tab:merged_words_nuggets_density_energy} reports nuggets/Wh across conditions, demonstrating small but significant absolute differences. Overall, \emph{llama} is the most efficient model with the mean nuggets/Wh ranging from $2.005$(IMP)$ - 2.574$ (PE), compared to $1.331$(IMP)$ -1.97$(PE) for \emph{deepseek}. 
Within-agent deltas (Table~\ref{tab:quick_all}) highlight two stable patterns:
ES vs HE: \(-14.0\%\) (deepseek), \(-23.2\%\) (llama), \(-11.1\%\) (qwen);
PE vs IMP: \(+48.7\%\) (deepseek), \(+26.0\%\) (llama), \(+37.1\%\) (qwen).
Thus, impoliteness is energy-inefficient; HE/PE deliver more nuggets per Wh.\\

\noindent\textbf{Omnibus Tests.} Likelihood-ratio $\chi^2$ tests show significant effects for Cluster: $\chi^2_{(4)}=2705.87$, $p$ < .001; AgentModel: $\chi^2_{(2)}=3345.16$, $p$ < .001; Cluster$\times$AgentModel: $\chi^2_{(8)}=303.43$, $p$ < .001. \\

\noindent\textbf{Pairwise exceptions (Tukey-adjusted).}
Again, most efficiency comparisons were significant, with only 6 non-significant pairs occurring, without any obvious trend.\\

\noindent{\emph{Note:}} Fig.~\ref{fig:h3_combined} shows a similar non-linearity for energy to that observed for information density in Section \ref{sec:H3a}. We see modest increases in Wh initially buy noticeably more nuggets, but growth slows and can plateau at higher energy budgets. This suggests that extra computation tends to lead to verbosity, formatting, or hedging rather than new factual units once the main answers to questions relating to a step have been covered. The differences between clusters thus reflect how efficiently they turn computations into nuggets, not merely how much energy they use.

%% file: 06_discussion.tex
\section{Discussion}

Our two-staged study offers a first look at how politeness strategies shape interactions with generative AI in the context of task-oriented information seeking dialogues. Annotating 30 cooking dialogues with an extended version of Leech's classification scheme for politeness-sensitive speech acts, we found users cluster along a spectrum from hyperpolite to hyperefficient, with intermediate styles reflecting distinct goals. Scaling up to 18k simulated dialogues, we show that politeness systematically influences reply length, informational yield, and energy use, suggesting that conversational style can have measurable effects on both content delivery and resource costs. Switching profiles can more than double the output length (+60–90\%) and increase information nuggets (+20–40\%), but at clear efficiency costs. Engagement-heavy styles give more but waste more, while impoliteness risks verbosity without value. The results also hint at diminishing returns: more words and watts eventually add little new information.

We begin by highlighting inclusion and environmental motivations for our work. Next, we revisit our results in light of these motivations and discuss their implications for the design of task-oriented information seeking, as well as directions for future work.

\subsection{Inclusion \& fairness}

\emph{Different styles, different rewards}. The different politeness strategies we observed resulted in consistent variation in outcomes. ES/HP users receive more informational units, but at higher costs in words and energy. HE users receive fewer units, but more concisely. While nugget counts do not equate to task utility, these systematic differences raise fairness concerns if certain user styles are consistently advantaged or disadvantaged. %resulted in consistent variation in outcomes. ES/PE users receive more nuggets, but pay higher costs in terms of words and energy. HE users, on the other hand, receive less information, but at a lower cost. The question is whether any of these groups is systematically disadvantaged for the same task. 
Politeness is culturally ingrained and varies according to demographic and situational factors. If systems do reward certain behaviours over others, this could amplify existing inequities.

\emph{Accessibility \& cognitive load.} Longer responses are not necessarily better. Even if they contain more information, users with time, attention, or bandwidth constraints may be penalised by verbosity.

\emph{Sensitivity to impoliteness.} 
In our synthetic impolite profile, we observe clear disadvantages, though
 some models' nugget counts are comparatively less sensitive to IMP prompts. %, even though reply length and efficiency do vary. 
 We consider this as a positive inclusion signal: users should not have to `pay' with politeness for access to information, particularly in high-stress or direct-speech contexts. At the same time, style-robustness and safety must co-exist: the aim is parity of informational help across everyday registers, not higher tolerance for harassment. Future work may wish to formalise a \emph{Style-Response Parity} metric (e.g., nugget parity ratios and efficiency parity across clusters) and test interventions that improve parity without inflating costs or weakening guardrails.

\subsection{Environmental \& resource costs}

\emph{Polite polluters.} We found significant differences in the energy costs of different politeness strategies, with the more engaging strategies often being less efficient and more energy-hungry. While the effect sizes in absolute terms are small, these will scale, and will undoubtedly be amplified by larger models.

\emph{Diminishing returns.} We observed consistent plateaus across models and politeness profiles, showing that additional words or energy provide ever smaller information gains. This suggests an opportunity to limited verbosity once the plateau begins, without reducing nuggets.

\subsection{Design \& Policy Implications}

We identify a number of aspects for system designers and service providers to consider based on our findings:

\emph{Parity across styles.} Systems should aim for a comparable information density between styles to avoid politeness becoming a hidden `tax' for access to detail.

\emph{Preference-aware generation.} Interfaces may allow users to set persistent preferences for response length or efficiency, ensuring that their chosen style, whether brief or elaborate, does not disadvantage them.

\emph{Explanatory transparency.} Compact, human-readable rationales (e.g., ``This reply is concise; click to expand'') may reduce pressure on users to adopt more engaging styles merely to obtain fuller answers.

\emph{Guardrails for impoliteness.} If impolite prompts tend to inflate reply length without adding value, policy should limit verbosity by summarising content and offering structured follow-ups. This could prevent inefficiency spirals while maintaining access to useful information.

%Consistent trends across models (even if magnitudes are different).

%Hyperpolite, although an extreme outlier in Frummet et al's data, did not result in extreme outcomes.

\subsection{Limitations}

Our study is limited to cooking conversations as an example of stepwise, task-oriented interaction. The behaviours we observed come from a small unrepresentative sample and the clusters we found were uneven in size. Generalisation, especially to other domains (e.g., learning, troubleshooting, health), remains to be validated. Moreover, information nuggets do not necessarily equate to utility. Future work should validate that nuggets correlate with task success, accuracy, and user satisfaction. Finally, for resource reasons, we only studied compact LLMs. Future work should aim to replicate with larger and multilingual models to test generalisability.

\section{Conclusions and Future work}
 %Our findings demonstrate that users' conversational choices, as well as model characteristics, shape both content and cost. This highlights important implications for system and policy design, but also points to open questions. A deeper understanding of user motivations, including social needs, response preferences, and cross-cultural variation, is needed to guide the development of GenAI information seeking systems that are both inclusive and sustainable.
We studied how user politeness profiles shape GenAI behaviour in task-oriented cooking dialogues, combining empirical analysis of human interactions with large-scale LLM--LLM simulation. 
Our findings demonstrate that users' conversational choices, alongside model characteristics, systematically affect response length, information transfer, and energy efficiency. 
This highlights important implications for system and policy design, suggesting that politeness can act as a hidden cost factor with consequences for both inclusion and sustainability. 
 A deeper understanding of user motivations, including social needs, response preferences, and cross-cultural variation, is needed to guide the development of GenAI information seeking systems that are both inclusive and sustainable.
\newpage